\documentclass{aa}
\usepackage{graphicx}
\begin{document}

\title{The WEBT \object{BL Lac} Campaign 2000} 


\author{M.~Villata \inst{1}
\and C.~M.~Raiteri \inst{1}
\and O.~M.~Kurtanidze \inst{2,3,4}
\and M.~G.~Nikolashvili \inst{2}
\and M.~A.~Ibrahimov \inst{5,6}
\and I.~E.~Papadakis \inst{7,8}
\and K.~Tsinganos \inst{7}
\and K.~Sadakane \inst{9}
\and N.~Okada \inst{9}
\and L.~O.~Takalo \inst{10}
\and A.~Sillanp\"a\"a \inst{10}
\and G.~Tosti \inst{11}
\and S.~Ciprini \inst{11}
\and A.~Frasca \inst{12}
\and E.~Marilli \inst{12}
\and R.~M.~Robb \inst{13}
\and J.~C.~Noble \inst{14}
\and S.~G.~Jorstad \inst{14,15}
\and V.~A.~Hagen-Thorn \inst{15,16}
\and V.~M.~Larionov \inst{15,16}
\and R.~Nesci \inst{17}
\and M.~Maesano \inst{17}
\and R.~D.~Schwartz \inst{18}
\and J.~Basler \inst{18}
\and P.~W.~Gorham \inst{19}
\and H.~Iwamatsu \inst{20}
\and T.~Kato \inst{20}
\and C.~Pullen \inst{21}
\and E.~Ben\'{\i}tez \inst{22}
\and J.~A.~de~Diego \inst{22}
\and M.~Moilanen \inst{23}
\and A.~Oksanen \inst{23}
\and D.~Rodriguez \inst{24}
\and A.~C.~Sadun \inst{25}
\and M.~Kelly \inst{25}
\and M.~T.~Carini \inst{26}
\and H.~R.~Miller \inst{27}
\and S.~Catalano \inst{12}
\and D.~Dultzin-Hacyan \inst{22} 
\and J.~H.~Fan \inst{28}
\and R.~Ishioka \inst{20}
\and H.~Karttunen \inst{10}
\and P.~Kein\"anen \inst{10}
\and N.~A.~Kudryavtseva \inst{15}
\and M.~Lainela \inst{10}
\and L.~Lanteri \inst{1}
\and E.~G.~Larionova \inst{15}
\and K.~Matsumoto \inst{20}
\and J.~R.~Mattox \inst{29}
\and F.~Montagni \inst{17}
\and G.~Nucciarelli \inst{11}
\and L.~Ostorero \inst{30}
\and J.~Papamastorakis \inst{7,8}
\and M.~Pasanen \inst{10}
\and G.~Sobrito \inst{1}
\and M.~Uemura \inst{20}
} 

\offprints{M.\ Villata, \email{villata@to.astro.it}} 

\institute{Istituto Nazionale di Astrofisica (INAF), Osservatorio Astronomico 
di Torino, Via Osservatorio 20, 10025 Pino Torinese (TO), Italy 
\and Abastumani Observatory, 383762 Abastumani, Georgia
\and Astrophysikalisches Institute Potsdam, An der Sternwarte 16, 14482 
Potsdam, Germany
\and Landessternwarte Heidelberg-K\"onigstuhl, K\"onigstuhl 12, 69117 
Heidelberg, Germany
\and Ulugh Beg Astronomical Institute, Academy of Sciences 
of Uzbekistan, 33 Astronomical Str., Tashkent 700052,
Uzbekistan
\and Isaac Newton Institute of Chile, Uzbekistan
Branch
\and Physics Department, University of Crete, 710 03 
Heraklion, Crete, Greece
\and IESL, Foundation for Research and Technology-Hellas, 
711 10 Heraklion, Crete, Greece
\and Astronomical Institute, Osaka Kyoiku University, 
Kashiwara-shi, Osaka, 582-8582 Japan
\and Tuorla Observatory, 21500 Piikki\"o, Finland
\and Osservatorio Astronomico, Universit\`a di Perugia, 
Via B.\ Bonfigli, 06126 Perugia, Italy
\and Osservatorio Astrofisico di Catania, Via S.\ Sofia 
78, 95123 Catania, Italy
\and Department of Physics and Astronomy, University 
of Victoria, PO Box 3055, Victoria, BC V8W 3P6, Canada
\and Institute for Astrophysical Research, Boston 
University, 725 Commonwealth Ave., Boston, MA 02215, USA
\and Astronomical Institute, St.-Petersburg
State University, Bibliotechnaya Pl.\ 2,
Petrodvoretz, 198504 St.-Petersburg, Russia
\and Isaac Newton Institute of Chile, St.-Petersburg Branch
\and Dipartimento di Fisica, Universit\`a La Sapienza,
Piazzale A.\ Moro 2, 00185 Roma, Italy
\and Department of Physics and Astronomy,
University of Missouri-St.\ Louis,
8001 Natural Bridge Road,
St.\ Louis, MO 63121, USA
\and Jet Propulsion Laboratory, California Institute of 
Technology, 4800 Oak Grove Drive, Pasadena, CA 91109, USA
\and Department of Astronomy, Faculty of Science, 
Kyoto University, Kyoto, Japan
\and Clarke and Coyote Astrophysical Observatory, 
PO Box 930, Wilton, CA 95693, USA
\and Instituto de Astronom\'{\i}a, UNAM, 
Apdo.\ Postal 70-264, 04510 M\'exico DF, Mexico 
\and Nyr\"ol\"a Observatory, Jyv\"askyl\"an Sirius ry, 
Kyllikinkatu 1, 40950 Jyv\"askyl\"a, Finland
\and Guadarrama Observatory, C/ San Pablo 5, 
Villalba 28409, Madrid, Spain
\and Department of Physics, University of Colorado 
at Denver, PO Box 173364, Denver, CO 80217-3364, USA
\and Department of Physics and Astronomy, Western 
Kentucky University, 1 Big Red Way, Bowling Green, KY 42104, USA
\and Department of Physics and Astronomy, Georgia 
State University, Atlanta, GA 30303, USA
\and Center for Astrophysics, Guangzhou University, 
Guangzhou 510400, China
\and Department of Chemistry, Physics, \&
Astronomy, Francis Marion University, PO Box 100547, Florence, SC
29501-0547, USA   
\and Dipartimento di Fisica Generale, Universit\`a di Torino, 
Via P.\ Giuria 1, 10125 Torino, Italy 
}

\date{Received; Accepted;}

\titlerunning{The WEBT BL Lac Campaign 2000}

\authorrunning{M.\ Villata et al.}

\abstract{see next page
}

\maketitle
\twocolumn[{\bf Abstract.} We present $UBVRI$ light curves of 
\object{BL Lacertae} from May 2000 to
January 2001, obtained by 24 telescopes in 11 countries. More than 15000 
observations were performed in that period, which was the extension
of the Whole Earth Blazar Telescope (WEBT) campaign
originally planned for July--August 2000. The exceptional sampling reached
allows one to follow the flux behaviour in fine details.  Two different phases can
be distinguished in the light curves: a first, relatively low-brightness phase
is followed by an outburst phase, after a more than $1\rm\,mag$ brightening in a
couple of weeks. Both the time duration (about $100\rm\,d$) and the
variation amplitude (roughly $0.9\rm\,mag$) are similar in the two phases. Rapid
flux oscillations are present all the time, involving variations up to a few
tenths of mag on hour time scales, and witnessing an intense intraday activity
of this source. In particular, a half-mag brightness decrease in about 
$7\rm\,h$ was detected on August 8--9, 2000, immediately 
followed by a $\sim 0.4 \rm \, mag$ brightening in $1.7\rm\,h$. 
Colour indexes have been derived by coupling the highest precision $B$ and $R$ data 
taken by the same instrument within $20\rm\,min$ and after subtracting the
host galaxy contribution from the fluxes. The 620 indexes obtained show that
the optical spectrum is weakly sensitive to the long-term trend, while it
strictly follows the short-term flux behaviour, becoming bluer when the
brightness increases. Thus, spectral changes are not related to the host galaxy
contribution, but they are an intrinsic feature of fast flares. We suggest that
the achromatic mechanism causing the long-term flux base-level modulation can
be envisaged in a variation of the relativistic Doppler beaming factor,
and that this variation is likely due to a change of the
viewing angle. Discrete correlation function (DCF) analysis reveals the existence of
a characteristic time scale of variability of $\sim 7\rm\,h$ in the light
curve of the core WEBT campaign, while no measurable time delay between
variations in the $B$ and $R$ bands is found.

\bigskip]

\section{Introduction}

\object{BL Lacertae} is a well-known source that has been observed in the optical
band for more than a century. It has been used to define a whole class of 
active galactic nuclei (AGNs), which are characterized by absence or extreme
weakness of the emission lines, intense variability at all wavelengths, high
polarization, and superluminal motion of radio components. The BL Lacertae objects,
together with the flat-spectrum radio quasars, are known as ``blazars''. 

Although the details of blazar emission are under debate, the commonly accepted
general scenario foresees a central black hole fed by an accretion disc,
and a plasma jet which is responsible for the non-thermal continuum. 
In order to explain several observational evidences, the emitted radiation
is assumed to be relativistically beamed towards us.
The low-energy emission, from the
radio band to the UV--X-ray region, is likely synchrotron radiation produced by
ultra-relativistic electrons in the jet. The origin of the higher-energy
radiation, up to $\gamma$-rays, is less clearly established: it is
reasonable to suppose that the soft radiation produced by the synchrotron
process can be inversely comptonized up to the $\gamma$-ray energies (SSC
models), but it is also possible that the photons to be comptonized come from out of
the jet, either directly from the accretion disc or from the broad line region
(EC models). Recent observations seem to indicate that a mixture of SSC and EC
processes may be at work in the blazar jets (see e.g.\ Madejski et al.\ \cite
{mad99} and B\"ottcher \& Bloom \cite{boe00} for the case of BL Lacertae). 
Another possibility is that the high-energy emission is produced by pair 
cascades coming from the interaction between soft photons and highly 
relativistic protons (proton models).

The violent flux variations observed in blazars
have been explained in a variety of ways: shocks travelling down
the jet (e.g.\ Marscher \cite{mars96}), changes of the
Doppler factor due to geometrical reasons (e.g.\ Dreissigacker \& Camenzind
\cite{dre96}; Villata \& Raiteri \cite{vil99}), accretion disc instabilities
(e.g.\ Wiita \cite{wii96}), gravitational microlensing
(e.g.\ Schneider \& Weiss \cite{sch87}). The observation of microvariability, that
is of flux changes on time scales of less than a day, raises the question of
what is the smallest time scale of variability in blazars and, if the
variations are of intrinsic nature, of how small the size of the emitting
region can be. 

Variability studies are thus a powerful tool to investigate blazar emission and
to discriminate among the various theoretical interpretations, in particular
when observations are done in a continuous way and simultaneously at different
wavelengths. This is why in the last years optical observers have set up
collaborations to make the observational effort more efficient. 

The Whole Earth Blazar Telescope (WEBT;
http:// www.to.astro.it/blazars/webt/) is an international
organization that includes about 30 observatories located all around the
world. Its aim is to obtain accurate and continuous monitoring of a source
during a time-limited campaign (from few days to several weeks), which is
often organized in concert with satellite observations in the X- and
$\gamma$-rays, and ground-based observations in the radio band and at TeV
energies. The location at different longitudes of its members allows them to
optimize observations during the 24 hours of the day, gaps due to daylight being, in
theory, extremely small. In practice, limitations due to bad weather conditions,
telescope overscheduling, technical problems are present, but in any case this
monitoring strategy has already demonstrated that it can provide unprecedentedly dense
sampling (see Villata et al.\ \cite{vil00} about the WEBT campaign on \object{S5
0716+71} of February 1999, and Raiteri et al.\ \cite{rai01} about the
first-light WEBT campaign on \object{AO 0235+16} of November 1997), and even better
results are expected by the robotization of at least some of the telescopes
participating in the WEBT.

In this paper we report on $UBVRI$ photometric monitoring of BL
Lacertae during the summer 2000 WEBT campaign and its extension (May 2000 --
January 2001). The core optical campaign took place simultaneously with the
planned high-energy campaign coordinated by M.\ B\"ottcher, involving X-ray
and TeV observatories such as BeppoSAX, RXTE, CAT, and HEGRA
(B\"ottcher et al.\ \cite{boe02}). 
Previous WEBT campaigns on BL Lacertae had been organized in June
1999, in conjunction with observations by the BeppoSAX and ASCA satellites
(Mattox \cite{matt99}).
The results of these campaigns are presented in Ravasio et al.\ 
(\cite{rav02}) and Villata et al.\ (\cite{vil02}).

The present paper is organized as follows: in Sect.\ 2 we review 
optical studies on BL Lacertae; the observing strategy and data
reduction/assembling procedures are described in Sect.\ 3. $UBVRI$ light
curves are presented in Sect.\ 4, colour indexes are discussed in Sect.\ 5, and the 
autocorrelation study can be found in Sect.\ 6. Discussion and conclusions are
drawn in Sect.\ 7.

\section{BL Lacertae}

The AGN BL Lacertae lies within an
elliptical galaxy (Miller et al.\ \cite{mil78}), at a distance $z=0.0688 \pm
0.0002$ (Miller \& Hawley \cite{mil77}). Ejection and evolution
of four highly-polarized superluminal radio components moving on curved
trajectories have been observed by Denn et al.\ (\cite{den00}) with the VLBA.

One puzzling feature is that, although BL Lacertae stands as the prototype of
a whole class of objects in which emission lines are absent or extremely weak,
on some occasions broad H$\alpha$ and H$\beta$ emission lines were
found in its spectrum, raising the issue of its membership to the class named after it
(Vermeulen et al.\ \cite{ver95}). Corbett et al.\ (\cite{cor00}) analysed eight
spectra taken over a period of 30 months, from June 1995 to December 1997, and
found that the equivalent width of the H$\alpha$ line varies approximately
inversely with the optical continuum flux: this suggests that the
broad-line region is likely not photoionized by the beamed synchrotron
radiation of the jet (which nevertheless cannot be ruled out), but by
radiation coming from the hot accretion disc.

BL Lacertae has been observed for more than a century in the optical band, and
it is well known for its intense variability on both long (months, years) and
short (days or less) time scales. 
Since the early work by Racine (\cite{rac70}), microvariability was
detected in optical observations of this source. Miller et al.\ (\cite{mil89})
observed a $0.12 \, \rm mag$ variation in $1.5\rm\,h$. 

Carini et al.\ (\cite{car92}) reported on 17 years of optical monitoring, in
which the source exhibited ``erratic'' behaviour, its $V$ magnitude varying
between 14.0 and 16.0. From their $B-V$ versus $V$ plot no
well-defined correlation between brightness and colour appeared, but a ``bluer
when brighter'' trend is recognizable. The authors searched for
microvariability, and found several episodes of variations as fast as $0.1 \,
\rm mag\,h^{-1}$, the most rapid rate of change observed being $0.19 \, \rm
mag\,h^{-1}$. 

$BVRI$ photometry of BL Lacertae in 1993--1995 was presented by Maesano et
al.\ (\cite{mae97}), confirming the spectral flattening for a flux increase.

A big observing effort was undertaken during the summer 1997 broad
optical outburst, announced by Noble et al.\ (\cite{nob97}). Subsequent
circulars reported on high flux levels also in the $\gamma$- and X-rays
(Hartman et al.\ \cite{har97}; Grove \& Johnson \cite{gro97}; Madejski et al.\
\cite{mad97}; Makino et al.\ \cite{mak97}).

Webb et al.\ (\cite{web98})
carried out $BVRI$ observations during the outburst, and found that variations were
simultaneous in all bands, but of higher amplitude at the higher frequencies,
and that there was a marginal evidence of a spectral flattening when the source
brightens. These two latter features were recognized also in the
microvariability events detected by Nesci et al.\ (\cite{nes98}) and Speziali 
\& Natali (\cite{spe98}), presenting variations up to $0.4\rm\,mag\,h^{-1}$.
$VR$ observations on 11 nights in July 1997 were performed by Clements \& Carini
(\cite{cle01}), who detected nightly variations from $0.1$ to $0.6 \, \rm mag$.
They also found that \object{BL Lac} became bluer when brighter and commented that
it is not clear whether the colour changes can be ascribed to the AGN or
are rather due to a greater contribution from the underlying galaxy when the
AGN is fainter.     
An extremely fast brightening of $0.6 \, \rm mag$ in $40\rm\,min$
was detected by Matsumoto et al.\ (\cite{mat99}) on August 2, 1997, inside 
a larger flux increase of more than $1\rm\,mag$ between 17 and 19 UT, confirmed 
by Ghosh et al.\ (\cite{gho00}). In the same night, the decreasing phase of the 
flare was observed by Massaro et al.\ (\cite{mas98}) and Speziali \& Natali 
(\cite{spe98}) as a variation of more than half a mag in about $2\rm\,h$ (see 
also Ghosh et al.\ \cite{gho00}).
Rapid and large-amplitude flux variations were also observed by Sobrito et
al.\ (\cite{sob99}) and Tosti et al.\ (\cite{tos99}). These papers also
contain extended light curves during the 1997 outburst.

During the
optical outburst, the EGRET instrument on the Compton Gamma Ray Observatory
revealed a $\gamma$-ray flux more than 4 times the previous detection, and a
harder spectrum than before (Bloom et al.\ \cite{blo97}). Moreover, a
noticeable $\gamma$-ray flux increase observed on July 18--19 apparently
preceded by several hours a brief optical flare. The RXTE and ASCA
satellites detected an X-ray flux respectively 2--3.5 times and more than 3
times higher than measured by ASCA in 1995 (Madejski et al.\ \cite{mad99};
Tanihata et al.\ \cite{tan00}). The multiwavelength spectrum of BL Lac during
the July 1997 outburst was examined by B\"ottcher \& Bloom (\cite{boe00}) in
terms of a homogeneous jet model. 

Intraday variability was found also the next year, in summer 1998, when the
source was in a fainter state (Massaro et al.\ \cite{mas99}; Nikolashvili et
al.\ \cite{nik99}).

$VRI$ photometry of BL Lacertae in 1997--1999 was presented by Fan et al.\
(\cite{fan01}): they detected microvariations with amplitude decreasing with
increasing wavelength. They also analysed the correlation among bands, finding no time
delay longer than $0.2\rm\,d$.

An analysis of colour variability of BL Lac during the 1997 and 1999 outbursts
was performed by Hagen-Thorn et al.\ (\cite{hag02}). They showed that in both
cases the spectral energy distribution remained unchanged during the outburst,
and that the spectrum was flatter in the more powerful outburst of 1997.

Since BL Lacertae is one of the few blazars for which time-extensive light
curves exist, a number of investigations have been devoted to the search for 
periodicities in its flux variations. 

Recurrent variations every 0.31, 0.60, and $0.88\rm\,yr$ were
recognized by Webb et al.\ (\cite{web88}) by means of a discrete Fourier
transform (DFT) analysis of their light curves, extending from June 1971 to January
1985. No evidence of periodicity was found instead by Carini et al.\
(\cite{car92}). 

``Whitening'' of time series was the method used by Marchenko et al.\
(\cite{mar96}; see also Hagen-Thorn et al.\ \cite{hag97}) to search for
periodicities in a 20-year long light curve of BL Lac: they found that only a
long-term component of $P = 7.8 \pm 0.2\rm\,yr$ is statistically
significant. 

In their study of the long-term optical base-level
fluctuations in AGNs, Smith \& Nair (\cite{smi95}) identified a best-fit,
well-defined cycle of $7.2\rm\,yr$ for the baseline meanderings shown by
the 20-year BL Lac light curve of the Rosemary Hill Observatory. It is
interesting to notice that the mere application of the Fourier analysis led
the authors to derive a period of $7.7\rm\,yr$, in fair agreement with
the results obtained by Marchenko et al.\ (\cite{mar96}).

Fan et al.\ (\cite{fan98}) analysed the historical optical light
curve of BL Lacertae with the Jurkevich method and derived a long-term period
of $\sim 14\rm\,yr$.

\section{Observations during the WEBT campaign}

Originally, the WEBT campaign was planned for the period July 17 -- August 11,
2000, that is from one week before to one week after the planned high-energy
campaign (B\"ottcher et al.\ \cite{boe02}). Simultaneous observations were
also performed in the radio band from the University of Michigan Radio
Astronomy Observatory (UMRAO) and from the Mets\"ahovi Radio Observatory, in
Finland. 

At the campaign start, BL Lacertae was in a relatively faint phase. A
subsequent, considerable brightening of the source in late September 2000
caused a campaign  extension up to January 2001.  

Data were finally collected from May 2000 to January 2001. 

Table 1 shows the list of observatories which participated in the WEBT
campaign. The name and country of the observatory (Column 1) is followed
by the telescope diameter (Column 2), by the total number of observations done
($N_{\rm obs}$, Column 3), and by the number of data points in $UBVRI$ derived, 
sometimes after binning (Columns 4--8).

\begin{table*} 
\caption{List of participating observatories by longitude} 
\begin{tabular}{lcrrrrrr} 
\hline 
Observatory & Tel.\ size (cm)& $N_{\rm obs}$ & $N_U$ & $N_B$ 
& $N_V$ & $N_R$ & $N_I$ \\ 
\hline  
Kyoto, Japan                      & 25 & 607  & 0 & 0  & 0 & 77  & 0 \\
Osaka Kyoiku, Japan               & 51 & 1209 & 0 & 0  &56 & 463 & 53 \\
Mt.\ Maidanak (AZT-22), Uzbekistan &150 &447   &59 &100 & 59&  131& 57 \\
Mt.\ Maidanak (T60-K), Uzbekistan  &60  &743   &99 &128 &133&  0  & 0  \\
Abastumani, Georgia (FSU)         &70  &2743  & 0 & 0  & 0 & 1253& 0   \\
Crimean, Ukraine                  &20  &2470  & 0 & 38 &67 &  159& 78  \\
Nyr\"ol\"a, Finland                   &40  &40    & 0 & 0  & 0 &  40 & 0   \\
Skinakas, Crete                   & 130& 630  & 0 & 313 & 0 & 314 & 0 \\
Catania, Italy                    & 91 & 1071 & 132 & 132 & 132 & 0 & 0 \\
Vallinfreda, Italy & 50 & 92 & 0 & 24 & 5 & 34 & 26 \\
Monte Porzio, Italy & 70 & 112 & 0 & 25 & 30 & 28 & 29 \\
Perugia, Italy & 40 & 673 & 0 & 0 & 76 & 527 & 63 \\
Torino, Italy & 105 & 900 & 0 & 83 & 57 & 578 & 54 \\
Guadarrama, Spain & 20 & 18 & 0 & 0 & 0 & 9 & 0 \\
Roque de los Muchachos (KVA), La Palma & 60 & 2018 & 0 & 148 & 1 & 401 & 0 \\
Roque de los Muchachos (NOT), La Palma & 256 & 74 & 0 & 7 & 0 & 58 & 9\\
Bell Farm, Kentucky & 60 & 1 & 0 & 0 & 0 & 1 & 0 \\
St.\ Louis, Missouri & 35 & 176 & 0 & 69 & 17 & 72 & 17 \\
Sommers-Bauch, Colorado & 60 & 8 & 0 & 0 & 3 & 3 & 0\\
Lowell, Arizona & 180 & 323 & 0 & 119 & 8 & 185 & 8 \\
San Pedro Martir, Mexico & 150 & 25 & 0 & 6 & 9 & 7 & 3\\
Palomar, California & 150 & 108 & 0 & 54 & 0 & 54 & 0 \\
Clarke and Coyote, California & 28 & 248 & 0 & 0 & 0 & 91 & 0 \\
University of Victoria, Canada & 50 & 889 & 0 & 0 & 0 & 363 & 0 \\
\hline
Total &  & 15625 & 290 & 1246 & 653 & 4848 & 397 \\
\hline
\end{tabular}

\medskip
$N_{\rm obs}$ is the total number of observations done, i.e.\ the number of
unbinned data; $N_U$, $N_B$, $N_V$, $N_R$, and $N_I$ are the 
numbers of data points in $UBVRI$ remained after discarding and binning some of the
original data
\end{table*}

\subsection{Observing strategy}

The participating observers were suggested to perform optical observations
alternately in two bands (Johnson's $B$ and Cousins' $R$, but other $R$
filters were also accepted) in order to obtain a $BR$ sequence of
frames throughout each observing night. 
Exposure times were chosen in view of a good compromise between high
precision and high temporal density. In those cases where high
precision would have required gaps of 15--$20\rm\,min$ in each light curve, 
it was suggested that
observations be obtained in the $R$ band only. As a matter of fact,
intensive $B$ monitoring was considered to be appropriate only for telescopes
larger than $1\rm\,m$. Moreover, at the beginning and end of the $BR$ (or $R$-only)
sequence, a complete set of filters ($U$)$BVRI$ (Johnson-Cousins when possible)
was suggested in order to follow the whole optical spectrum behaviour of the
source.

\subsection{Data reduction and assembling}

Data were collected as instrumental magnitudes of
the source and reference stars, in order to apply the same analysis and
calibration procedures to all datasets. Frame pre-reduction, i.e.\ bias/dark
correction and flat-fielding, was performed by the observers. 
Instrumental magnitudes were obtained by either standard aperture photometry
procedures or Gaussian fitting, in most cases by the observers themselves.

A minority of data came from photometer observations: these data were provided 
directly as standard magnitudes 

A careful data assembling was required, since the whole
dataset consisted of more than 15000 observations coming from 22 different
observatories, 24 different telescopes, and 25 different detectors. 

The analysis was performed through the following steps:  
\begin{itemize}
\item Obtaining standard
magnitudes of BL Lac from instrumental ones. Stars B C H K originally
identified by Bertaud et al.\ (\cite{ber69}; their b c h k) were used as
reference stars for the magnitude calibration of the source. The photometry
adopted was that published by Bertaud et al.\ (\cite{ber69}) for the $U$ and
$B$ bands, and the one from Fiorucci \& Tosti (\cite{fio96}) for the $VRI$
bands\footnote{In the error calculation, calibration errors were not taken
into account, since the main goal was to analyse flux variations, rather than 
to obtain a precise evaluation of the magnitude.}.    
\item Discarding bad and unreliable points.  
\item Binning the data when needed. 
\item Check of the final light curves for each observatory (see Table 1 for 
the relevant data numbers). 
\item Putting together different datasets, night by night. 
\item Estimating the offsets existing among different datasets (see Table
2) from overlapping data, and correcting them, night by night. 
\item Discarding less precise data in favour of higher-precision simultaneous data. 
\end{itemize} 

\begin{table*} 
\caption{Offsets given to the datasets in order to remove systematic 
differences; when a range is given, it means that different values 
were used in different nights; the accuracy of the offsets is similar 
to that of the corresponding datasets, ranging from less than 0.01 
to about $0.05\rm\,mag$; in few cases no comparison was possible 
and no offset was given} 
\begin{tabular}{l c c c c c} 
\hline 
Observatory & $U$ & $B$ & $V$ & $R$ & $I$ \\ 
\hline  
Kyoto & -- & -- & -- & 0 & -- \\
Osaka Kyoiku & -- & -- & $-0.05$ & $-0.04$--0 & 0 \\
Mt.\ Maidanak (AZT-22; SITe) & 0 & 0 & 0 & 0 & 0 \\
Mt.\ Maidanak (AZT-22; ST-7) & -- & $-0.27$--$-0.25$ & $-0.10$ &
$-0.21$--$-0.11$ & $-0.10$ \\  
Mt.\ Maidanak (T60-K) & $-0.15$ & $+0.10$ & $+0.10$ & -- & -- \\
Abastumani & -- & -- & -- & $-0.04$--$+0.06$ & -- \\
Crimean & -- & $-0.20$ & $-0.05$ & $-0.07$--$-0.04$ & $-0.05$ \\
Nyr\"ol\"a & -- & -- & -- & $+0.01$ & -- \\
Skinakas & -- & 0 & -- & $+0.02$ & -- \\
Catania & 0 & $+0.05$--$+0.15$ & $+0.05$ & -- & -- \\
Vallinfreda & -- & $+0.10$ & $-0.05$ & 0 & $-0.05$ \\
Monte Porzio & -- & $+0.10$ & $-0.05$ & $-0.07$ & $-0.05$ \\
Perugia & -- & -- & 0 & 0 & 0 \\
Torino & -- & $-0.13$--$-0.04$ & $-0.10$ & $-0.13$--$-0.01$ & $-0.10$ \\
Guadarrama & -- & -- & -- & 0 & -- \\
Roque de los Muchachos (KVA) & -- & 0 & 0 & $+0.02$ & -- \\
Roque de los Muchachos (NOT) & -- & $+0.05$ & -- & $-0.01$ & $-0.05$ \\
Bell Farm & -- & -- & -- & 0 & -- \\
St.\ Louis & -- & 0 & 0 & $+0.01$--$+0.02$ & 0 \\
Sommers-Bauch & -- & -- & 0 & $+0.04$ & -- \\
Lowell & -- & $+0.06$ & 0 & 0 & 0 \\
San Pedro Martir & -- & $-0.10$ & $-0.10$ & $-0.10$ & $-0.10$ \\
Palomar & -- & $-0.05$ & -- & $-0.125$ & -- \\
Clarke and Coyote & -- & -- & -- & $-0.04$ & -- \\
University of Victoria & -- & -- & -- & $+0.04$--$+0.11$ & -- \\
\hline
\end{tabular}
\end{table*}

Table 3 shows the number
of data points remained in the various bands at the end of the data assembling
procedure (to be compared with the totals 
of Table 1, where the last step of the assembling procedure is not considered), 
and the minimum and maximum magnitudes reached. 

\begin{table}
\caption{Number of points plotted in the $UBVRI$ light curves;
minimum and maximum magnitudes reached} 
\begin{tabular}{cccccc}
\hline
           & $U$ &  $B$ & $V$ & $R$  & $I$ \\
\hline
$N_{\rm points}$ & 290 & 1094 & 611 & 4015 & 396 \\
Min        & 14.49 & 14.76 & 13.75 & 13.01 & 12.35 \\
Max        & 16.08 & 16.26 & 15.21 & 14.50 & 13.74 \\
\hline
\end{tabular}
\end{table}

\section{\boldmath $UBVRI$ light curves}

The $UBVRI$ light curves of BL Lacertae from May 2000 to January 2001 are
presented in Fig.\ \ref{ubvri}. It is possible to identify two
well-defined phases: the first one starts with a  moderate
brightness increase, and sees the source in a relatively
low brightness state; it ends after one hundred days, when the brightness drops
to the minimum value ($\rm JD=2451788$, $R=14.50$). After that, a rapid
brightening of more than $1\rm\,mag$ in a couple of
weeks leads to the outburst phase, lasting again about one hundred days.
The last available observations witness a final dimming to the
pre-outburst levels.

   \begin{figure*}
   \centering
   \caption{$UBVRI$ light curves of BL Lacertae from May 2000 to January 2001;
   the numbers on the right are the numbers of data points in each light curve; the grey 
   (yellow in the electronic version) strip corresponds to the period of the core WEBT
   campaign}
   \label{ubvri}
   \end{figure*}

In both the two mentioned phases, flux oscillations of several tenths of
mag on day time scales are present, and it is interesting to notice
that the total variation amplitude is approximately the same (about $0.9\rm\,mag$)
in the pre-outburst and in the outburst periods, similar to the results reported 
for the 1997 outburst of BL Lac (Miller \cite{mil99}).

One striking feature of Fig.\ \ref{ubvri} is the exceptional
sampling obtained during the core WEBT campaign of July--August 2000, especially 
in the $R$ band.

The details of the brightness behaviour in the $R$ band during the core
WEBT campaign are visible in Fig.\ \ref{rtotcamp}. The most spectacular
variability was detected (with high precision and very good intranight
sampling) during the last part of the campaign, plotted in the lower panel of
Fig.\ \ref{rtotcamp}. The boxes indicate periods of particularly interesting
variations, which are shown in subsequent figures.

   \begin{figure*}
   \centering
   \caption{Light curve in the $R$ band during the core WEBT campaign}
   \label{rtotcamp}
   \end{figure*}

\begin{figure*}
   \sidecaption
   \includegraphics[width=13cm]{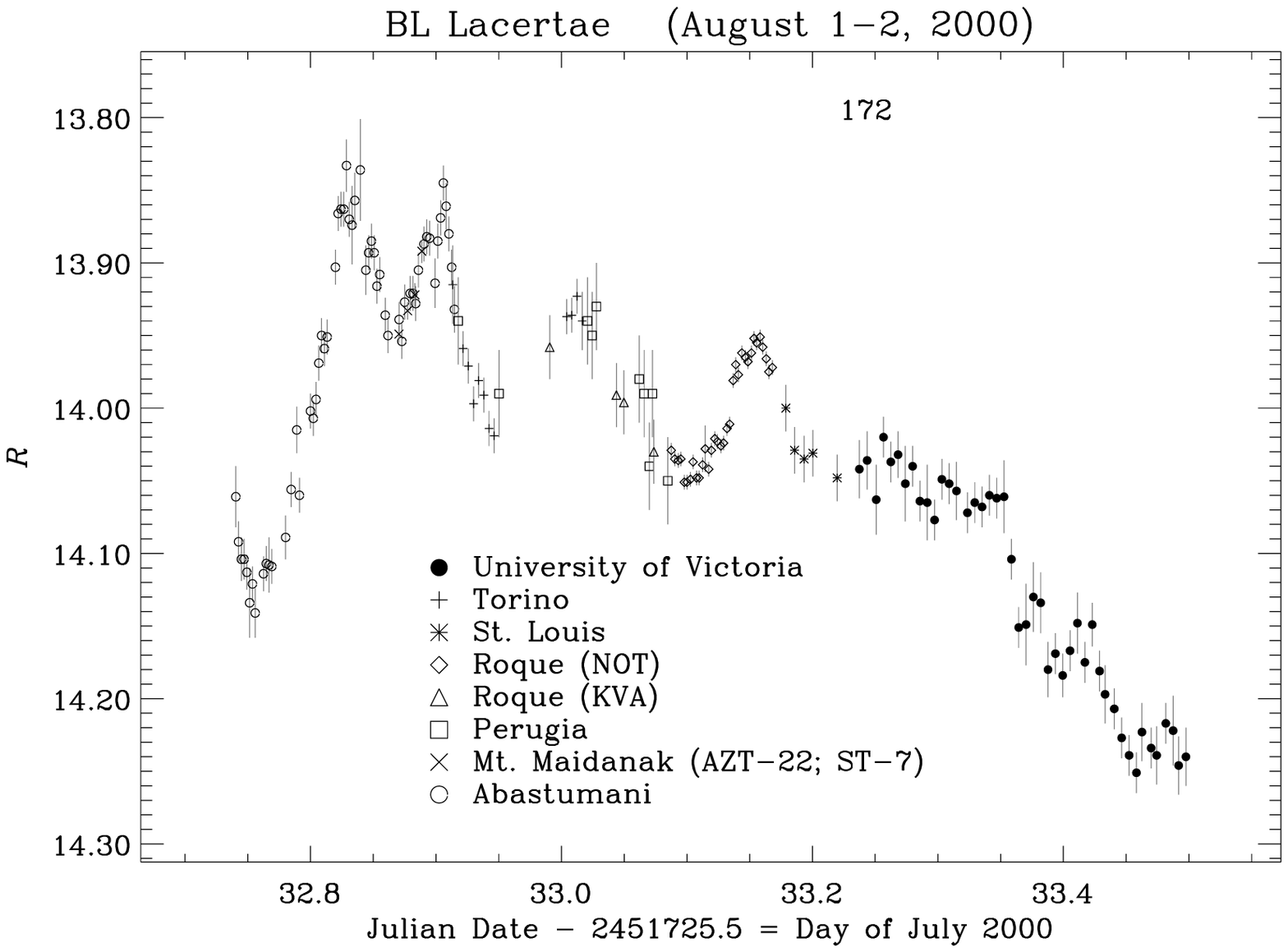}
   \caption{\ Light curve of BL Lacertae in the $R$ band on August 1--2, 2000}
   \label{rtot32-33}
   \end{figure*}

\begin{figure*}
   \sidecaption
   \includegraphics[width=13cm]{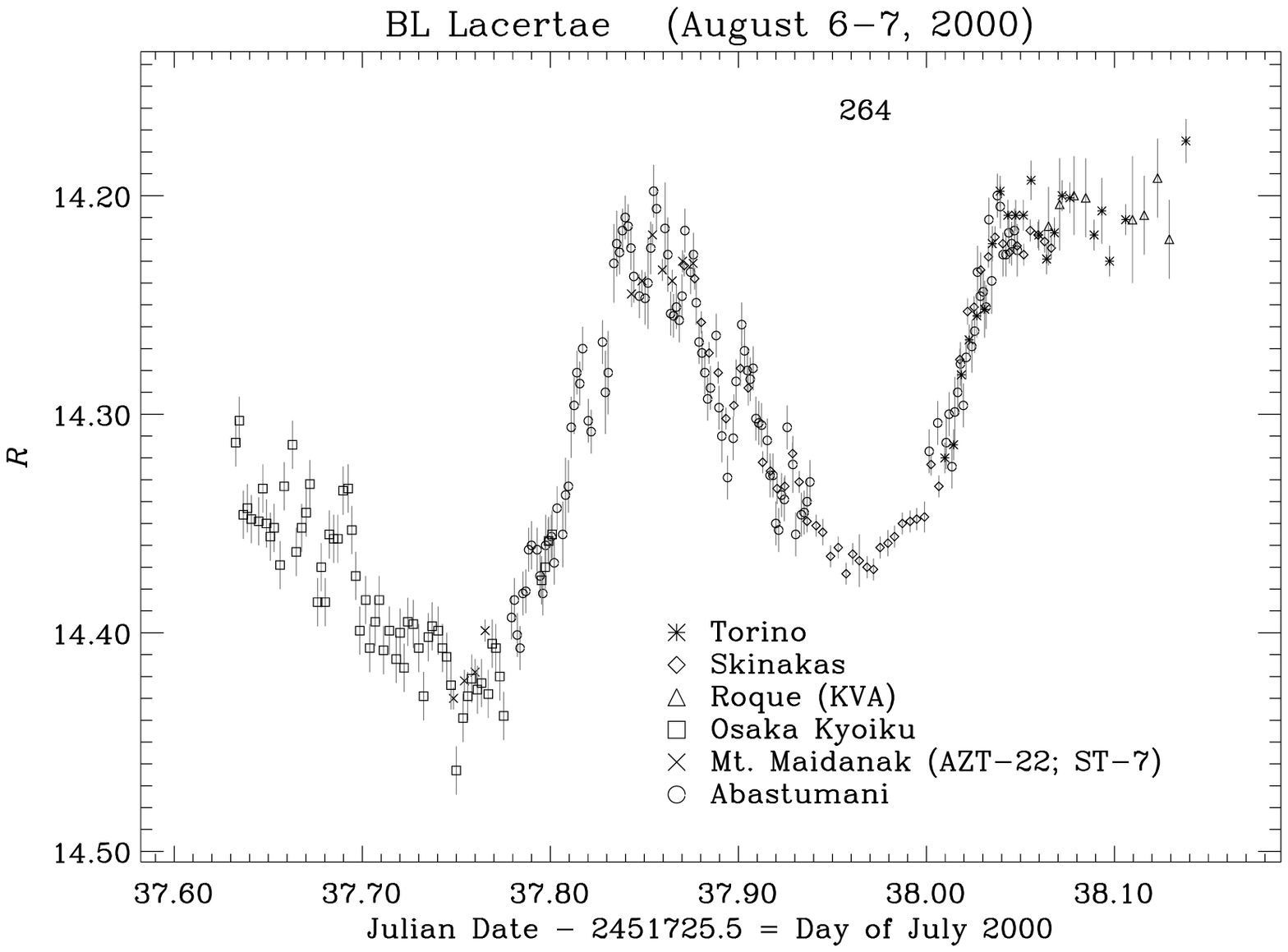}
   \caption{\ Light curve of BL Lacertae in the $R$ band on August 6--7, 2000}
   \label{rtot37-38}
   \end{figure*}

\begin{figure*}
   \sidecaption
   \includegraphics[width=13cm]{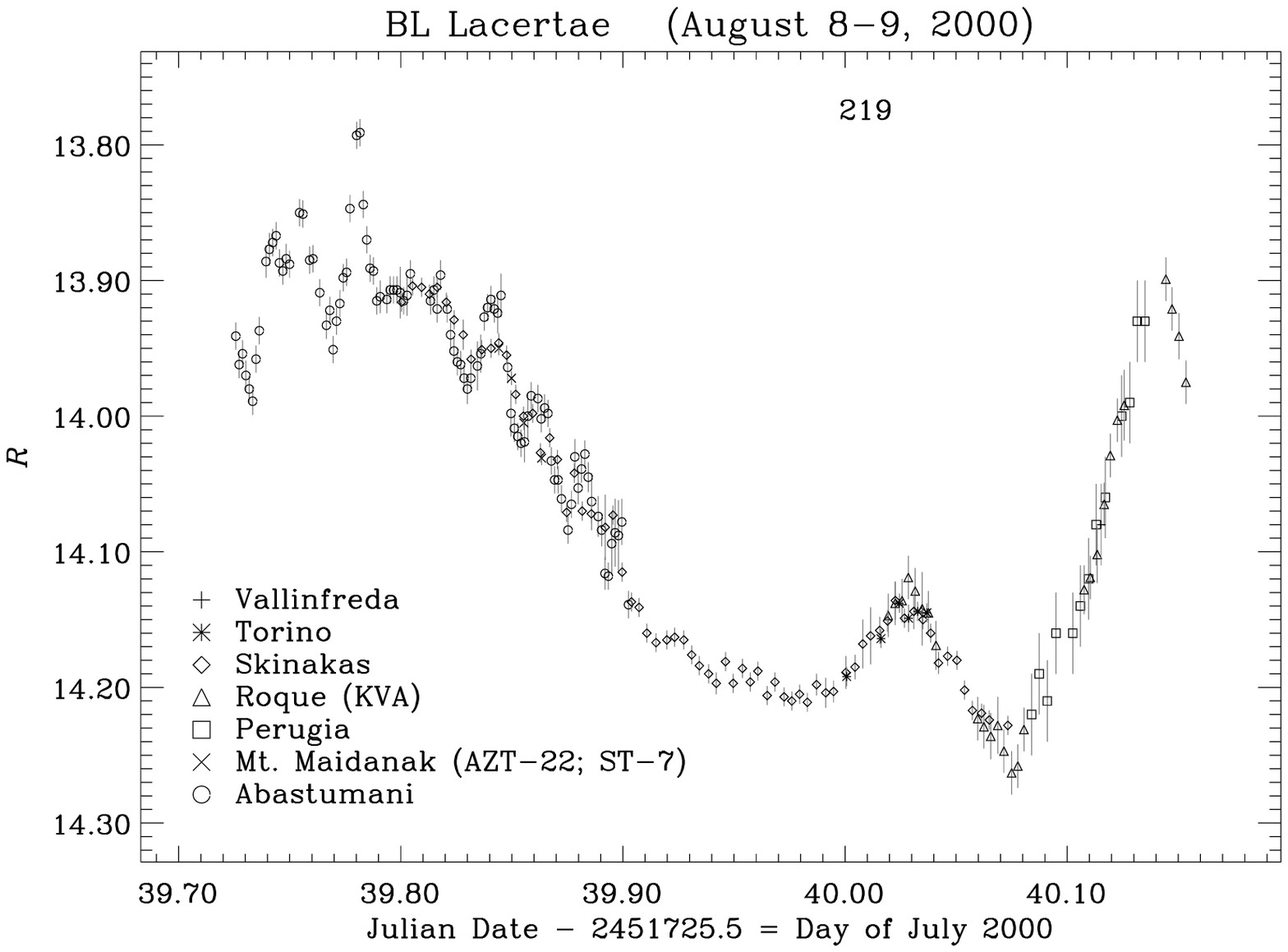}
   \caption{\ Light curve of BL Lacertae in the $R$ band on August 8--9, 2000}
   \label{rtot39-40}
   \end{figure*}

Data taken on August 1--2, August 6--7, and August 8--9, 2000 are
plotted in Figs.\ \ref{rtot32-33}, \ref{rtot37-38}, and \ref{rtot39-40},
respectively. The participating observatories are distinguished by
different symbols demonstrating how the observing task moves from east to
west as the Earth rotates. 

One can notice that, in principle, it is
possible to obtain a full-day observing coverage, when both the most eastern
and western observatories work, thus succeeding in filling the ``Pacific gap''.
In practice, this gap is often present in the WEBT light curve, but it is
limited to a few hours, which allows one to follow the brightness behaviour almost 
continuously.

The most impressive variations were detected on August 8--9: a half-mag brightness 
fall in about $7\rm\,h$, immediately followed by a very
steep brightening of $\sim 0.4 \rm \, mag$ in $1.7\rm\,h$.

By looking at the individual flaring episodes in the light curve of Fig.\ 
\ref{rtotcamp}, we 
can see that the most significant rising branches (on July
19, 23, 31, August 1, 3, 6, 7, 9) present brightening rates ranging from
0.11 to $0.22 \, \rm mag \, h^{-1}$. Moreover, one can notice that, in
general, rising slopes are steeper than dimming ones. However, since the
observed events might be the result of the superposition of different ``single''
flares, any speculation about the time
scales of the underlying brightening and dimming mechanisms is likely not very 
significant.

\section{Colour indexes}

We have seen that the brightness behaviour of BL Lacertae in the period examined
here appears as the superposition of rapid flares, typically lasting for
about a day or less, on a long-term trend, which is responsible for a
transition from a relatively low brightness level to a higher brightness level
around $\rm JD=2451800$. It is now interesting to analyse the data in order to
understand if the mechanism causing rapid flares is of the same nature as the
one determining the long-term flux variations.
One piece of information is likely to come from colour index analysis, which
can reveal whether flux variations imply spectral changes or not.

Since the BL Lacertae host galaxy is relatively bright, we first subtracted
its contribution from the observed fluxes in order to avoid contamination in the 
colour indexes.

According to Scarpa et al.\ (\cite{sca00}), the $R$ magnitude of the BL
Lacertae host galaxy is $R_{\rm host}=15.55 \pm 0.02$, adopting a host galaxy
colour  $V-R=0.61$; Mannucci et al.\ (\cite {man01}) derived an average
effective colour for elliptical galaxies with $M_V<-21$ of $B-V=0.99 \pm 0.05$.
The inferred $B$-band host galaxy magnitude is thus $B_{\rm host}=17.15$.

Actually, the above magnitudes represent extrapolations to infinity, while 
an aperture radius of $8 \rm \, arcsec$ (to be compared with the
$4.8 \rm \, arcsec$ half-light galaxy radius given by Scarpa et al.\
\cite{sca00}) was suggested in the data reduction for the source measure,
together with radii of 10 and $16 \rm \, arcsec$ for the edges of the
background annulus. By using these
parameters and a de Vaucouleurs $r^{1/4}$ profile, we estimated that the
host galaxy contribution to the observed fluxes is only 59.65\% of the whole
galaxy flux.

We then
transformed both the observed and the host galaxy $B$ and $R$ magnitudes
into dereddened fluxes by using the coefficient of Galactic extinction in 
the $B$ band $A_B=1.420$
given by NED and by deriving extinction in the Cousins' $R$ band by means of
Cardelli et al.\ (\cite{car89}): $A_R=0.9038$. Fluxes relative to zero-mag 
values were taken
from the photometric calibration of Bessell (\cite{bes79}).
The $B$ and $R$ host galaxy fluxes ($2.175 \, \rm mJy$ and $4.266 \, \rm mJy$,
respectively) were reduced by a factor 0.5965 and then subtracted from the
observed fluxes, and the point source (reddened) magnitudes derived from the
``cleaned'' fluxes.

$B-R$ colour
indexes were calculated by coupling $B$ and $R$ data taken by the same
instrument within $20\rm\,min$ (in most cases the time separation between 
the coupled data is in the range 2--$4\rm\,min$). Only $B$ and $R$ data 
with errors not greater
than 0.04 and $0.03\rm\,mag$, respectively, were considered. 

\begin{figure*}
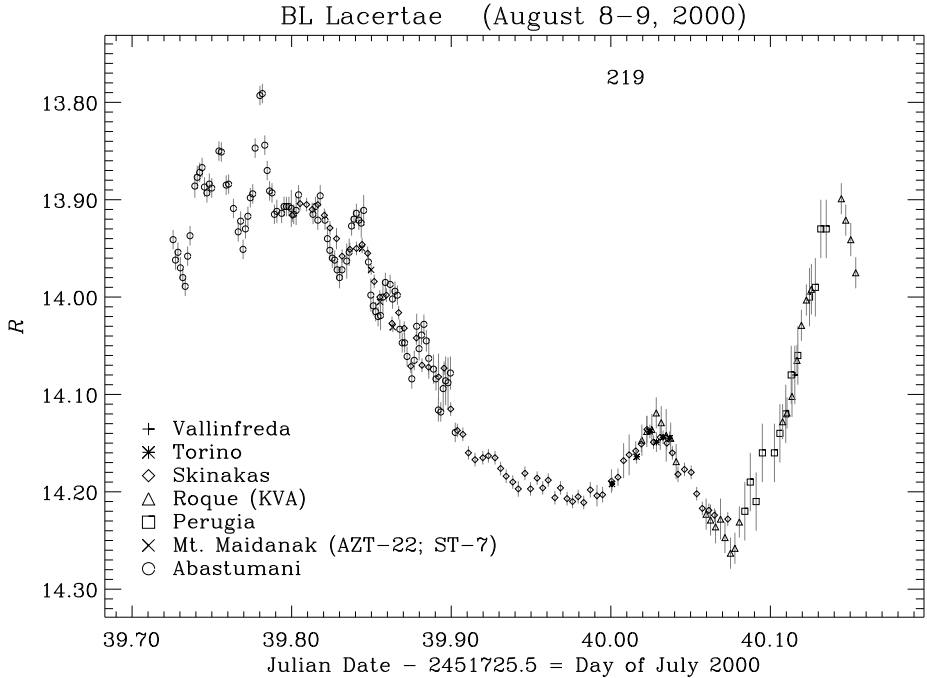
    
\sidecaption  
\caption{\ Temporal evolution of $B-R$ colour index (upper panel) and $R$
magnitude (lower panel) after subtracting the host galaxy contribution from
the fluxes}   
\label{colori_jd}          
\end{figure*}

The plot of the resulting 620 $B-R$ indexes as a function of time (see Fig.\
\ref{colori_jd}) suggests that colours are more sensitive to rapid variations
than to the long-term trend. 

\begin{figure*}    
\sidecaption  
\includegraphics[width=13cm]{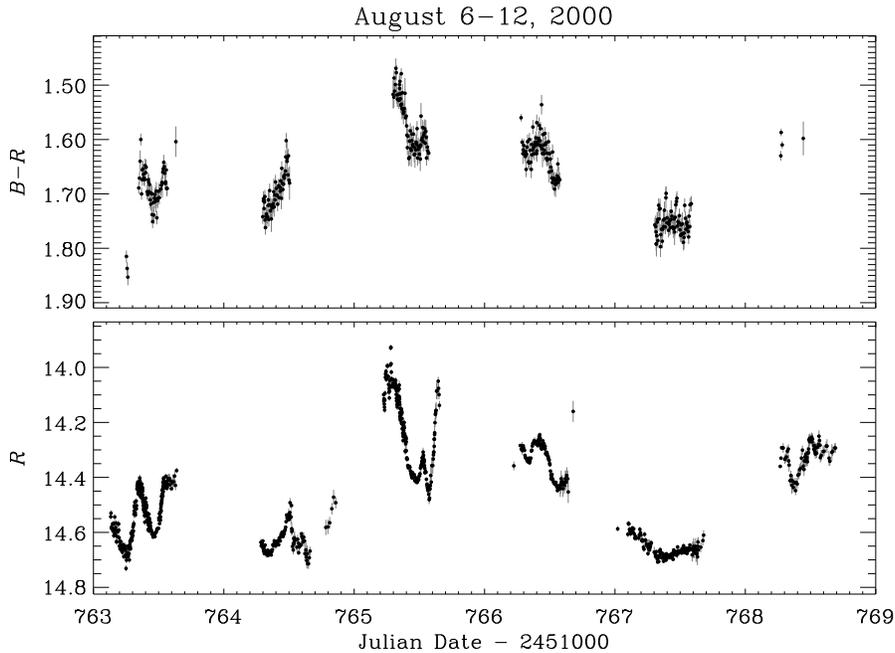}    
\caption{\ Temporal evolution of $B-R$ colour index (upper panel) and
$R$ magnitude (lower panel) during the last week of the core WEBT campaign,
after subtracting the host galaxy contribution from the fluxes}        
\label{colori_webt}         
\end{figure*}

Moreover, during well sampled flares the $B-R$ index strictly follows the
flux behaviour, as shown by Fig.\ \ref{colori_webt}, which presents an
enlargement of Fig.\ \ref{colori_jd} during the fourth week of the core WEBT
campaign. In this sense, we can say that fast flares are due to a chromatic
mechanism, which causes a spectral flattening when the source brightens.

\begin{figure*}    
\centering   
\includegraphics{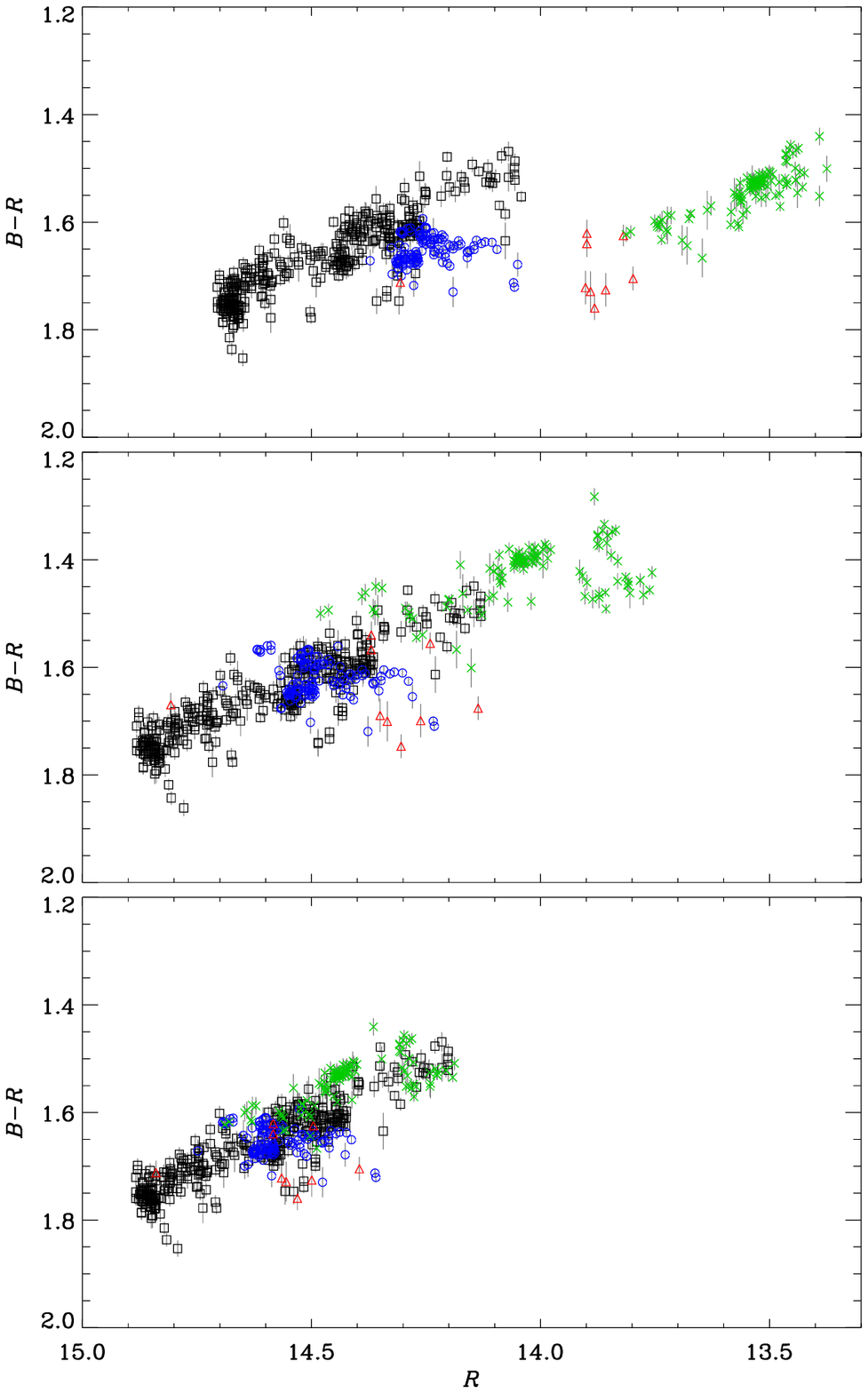}    
\caption{$B-R$ colour index versus $R$ magnitude (after subtracting the host
galaxy contribution from the fluxes) for uncorrected data (upper panel), for
data corrected by subtracting the base-level modulations modelled as a cubic
spline (middle panel), and for data further corrected for Doppler factor 
variations (lower panel); different symbols refer to different flux levels of
the spline (see Fig.\ \ref{spline}): $< 10 \, \rm mJy$ (black squares),
between  10 and $12.5\rm\,mJy$
(blue circles), between 12.5 and $15\rm\,mJy$ (red triangles), and $> 15 \, \rm
mJy$ (green crosses)}           
\label{b-r}        
\end{figure*}

Figure \ref{b-r} (upper panel) shows the $B-R$ versus $R$ plot. 
Points are distributed over two separated regions of the figure, according to the
brightness level of the source, with a boundary at $R\sim 14.1$. However,
inside each region, the colour indexes seem to follow a trend with a similar
slope: a bluer-when-brighter behaviour, as already noticed in Fig.\ 
\ref{colori_webt} for the short-term flares. On the contrary, the long-term
variations appear as essentially achromatic.

In order to verify the supposed existence of two different mechanisms acting
on different time scales, we have tried to model the long-term trend as a
modulated base contribution to the source flux density, on which the
short-term flares are superposed. We expect that, once fluxes are corrected for 
this contribution, the $B-R$ versus $R$ plot will contain the signature of one
component only, i.e.\ the chromatic one.

The first step has been to define a flux base level lapping on the flux minima
of the $R$ light curve.

Data in the $R$ band were first binned daily for removing effects due
to intranight dense sampling, and then binned over $10\rm\,d$; 
the binned light curve was then fitted with a cubic spline interpolation
(Press et al.\ \cite{pre92}). The $R$ light curve was divided into two
zones, corresponding to the pre-outburst ($\rm JD<2451790$) and outburst phases.
The previously derived spline was then proportionally rescaled to pass through
the minima of each zone.
The result can be seen in Fig.\ \ref{spline},
where the upper, grey (green in the electronic version) line traces the
original cubic spline interpolation, while the lower, dark (blue) line shows
the rescaled spline, representing the pursued base-level modulation
due to the achromatic mechanism mentioned above\footnote{A $10\rm\,d$ 
smoothing time scale appeared as the most suitable choice: a shorter 
one would have been too sampling dependent; a longer one would have 
smoothed too much the evident long-term variations which are visible 
in the best-sampled periods. In any case, changes of the binning 
interval in the range 5--$30\rm\,d$ do not affect the final results 
sensitively.}.

   \begin{figure*}
   \centering
   \caption{$R$-band fluxes (mJy) after subtraction of the host galaxy
contribution as a function of JD; cubic spline interpolation through the
binned light curve is represented by the grey (green) line; the rescaled 
spline (see text for explanation) is shown by the
dark (blue) line}        
   \label{spline}    
   \end{figure*}

The same spline was used to find the base-level modulation for
the $B$-band fluxes, by proportionally rescaling it to pass through the
minimum flux. The resulting flux ratio of the base levels is $F_R/F_B=2.332$
($B-R=1.787$). 

By seeing Fig.\ \ref{spline}, one might object that only a few minimum points
are very close to the base level, even when the sampling is good and many
local minima could be identified. The point is that local minima are most
likely not states where the flaring activity is out: they are presumably due to
the superposition of different events started at different times. On the
contrary, the detection of ``no flaring'' may be a very rare event, if ever
happens.

Fluxes were then ``corrected'' for their respective base levels
by subtracting the shaded (yellow) area shown in Fig.\ \ref{spline} from the
$R$ fluxes, and analogously for the $B$ fluxes.
``Corrected'' $B-R$ and $R$ values were finally
obtained by $B-R = 2.5 \log{(F_R/F_B)} + 0.868$, $R=-2.5 \log{F_R} + 17.125$,
where the constants take both the zero-mag fluxes and the
Galactic extinction coefficients into account. 

The new $B-R$ versus $R$ plot is shown in Fig.\ \ref{b-r} (middle panel): apart
from a few points mainly coming from the low-brightness peak of the spline
(red triangles, see caption to the figure), most of the points are now following a 
single linear trend,
confirming the starting assumption that an achromatic mechanism produces the
base-level variations.

However, one can notice that the distribution of data points corresponding to
the outburst phase [grey (green) crosses] still extends to higher brightness
levels. This is the consequence of the greater amplitude exhibited by the flux
variations in the outburst state (see Fig.\ \ref{spline}).
We have already noticed that, in the logarithmic scale of magnitudes,
variation amplitudes are comparable in the pre-outburst and in the outburst
phases, which means that flux amplitudes are proportional to the
flux level.

We can thus further refine our model for the flux base-level variations, by
assuming that the achromatic mechanism is also responsible for the
brightness-dependence of the variation amplitudes. A simple explanation for
this is obtained by assuming that the base-level oscillations are the 
result of the variation of the relativistic Doppler factor
$\delta=[\Gamma(1-\beta \cos\theta)]^{-1}$, where $\Gamma$ is the Lorentz
factor of the bulk motion of the plasma in the jet and $\theta$ is the
viewing angle, since fluxes are enhanced proportionally to a certain power of
$\delta$ by relativistic boosting. 

In order to clean the observed fluxes for this effect, we derived
``corrected'' fluxes by rescaling each original flux by the ratio between
the minimum value of the spline and the value of the spline at the considered
time. In this way, we obtain fluxes normalized to the value of $\delta$ where
the spline has its minimum, thus eliminating the effects of the $\delta$
variation, in terms of both the base-level variations and the different variation 
amplitudes. The resulting ``cleaned'' light curve is shown in Fig.\
\ref{flussi_cor}: the variation amplitude is now comparable over all the
period, which should mean that we are now seeing the behaviour of the
chromatic component alone.

\begin{figure}      
\caption{$R$-band fluxes after subtraction of the host galaxy
contribution and correction for the mechanism responsible for the
base-level variations (see text for explanation)}                
   \label{flussi_cor}    
   \end{figure}

As for the colour indexes, we derived $B-R$ and $R$ values from corrected 
fluxes as already done
in the previous case; the bottom panel of Fig.\ \ref{b-r} displays the final
result: all data fall in a narrower brightness range, as expected, and the
linear correlation appears better defined.

\section{Time correlations}

In the Introduction we mentioned a number of studies devoted to the search for
periodicities in the BL Lacertae light curves. 
The derived periods range from 0.31 to $14\rm\,yr$.
The light curves obtained by the WEBT collaboration extend to $\sim 240\rm\,d$,
so that possible periodicities longer than, say, a couple of months cannot be
looked for. On the other hand, the dense sampling allows one to test the existence
of characteristic times of variability down to very short periods.

\begin{figure}      
\resizebox{\hsize}{!}{\includegraphics{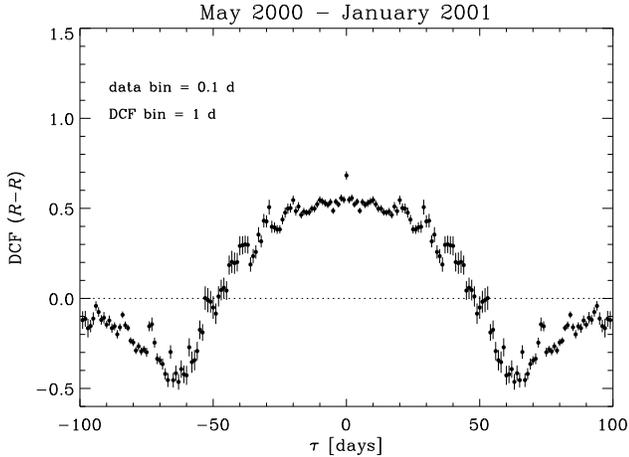}}    
\caption{Autocorrelation function for the uncorrected, galaxy-subtracted
$R$-band fluxes} 
\label{dcf_tot}        
\end{figure}

We applied the discrete correlation function (DCF) method (Edelson \& Krolik
\cite{ede88}; Hufnagel \& Bregman \cite{huf92}) to the BL Lacertae fluxes,
paying attention to the edge effects, as warned by Peterson (\cite{pet01}).
Figure \ref{dcf_tot} shows the autocorrelation of the uncorrected,
galaxy-subtracted $R$-band fluxes, with a $100\rm\,d$ maximum lag. The central,
wide maximum tells us that the $R$ light curve continues to correlate with
itself for time shifts shorter than a month; for larger time lags the
pre-outburst phase starts to overlap significantly with the outburst one, and
the DCF drops to negative values, implying anticorrelation. No significant
feature appears, which means that no reliable characteristic variability time
scale is found.

\begin{figure}     
\resizebox{\hsize}{!}{\includegraphics{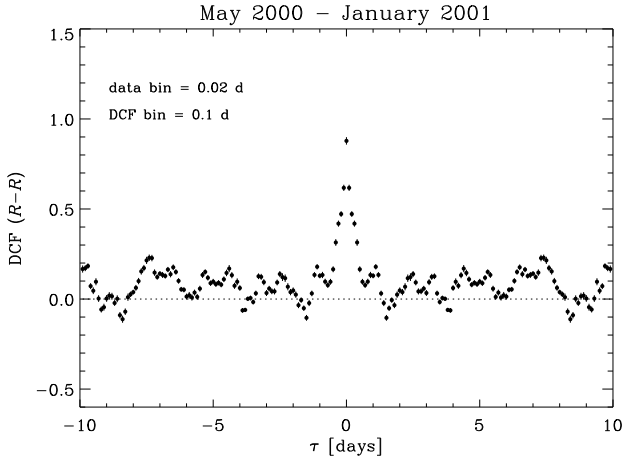}}    
\caption{Autocorrelation function for the $R$-band fluxes after subtracting
the host galaxy contribution and correcting for Doppler
factor variation}         
\label{dcf_cor}        
\end{figure}

Correction of fluxes for variation of the Doppler factor as
described in the previous section removes the signature of the long-term trend.
As one can see in Fig.\ \ref{dcf_cor}, where the autocorrelation  
for the corrected $R$-band fluxes is presented (zoomed
on smaller time lags), also in
this case no significant time scale is recognizable.

\begin{figure}     
\resizebox{\hsize}{!}{\includegraphics{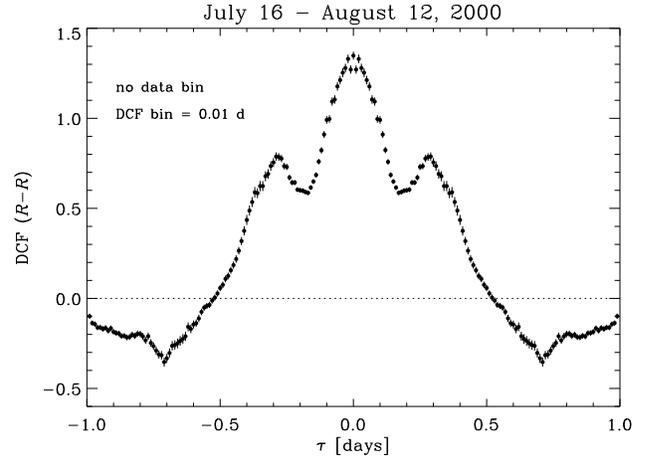}}    
\caption{The same as Fig.\ \ref{dcf_cor}, but restricted to the core WEBT campaign 
and zoomed on short time scales}
\label{dcf_zoom}        
\end{figure}

However, if we restrict the DCF analysis to the data from the core WEBT
campaign (see Fig.\ \ref{rtotcamp}), we see that a not negligible signal comes
out at a $\sim 7\rm\,h$ time scale, as shown by Fig.\ \ref{dcf_zoom}. This
feature is particularly evident in the second and fourth weeks of the core
campaign. Nevertheless, one has to notice that its significance might be 
affected by the lack of information during observing gaps.

In the above analysis we have neglected that not only fluxes, but also time
scales are affected by Doppler boosting. A reliable temporal analysis should
take this into account. 

It is known that time intervals are changed by a factor $\delta^{-1}$ by
relativistic Doppler effect. Under the assumption that fluxes $F$ are modified
proportionally to $\delta^3$, we have that
$\delta(t) / \delta_{\rm min} = \left[ F(t)/F_{\rm min} \right] ^{1/3} = \left[
s(t) / s_{\rm min} \right] ^{1/3}$, where $s(t)$ is the value of the spline
representing the flux base-level variations at time $t$, and $s_{\rm min}$ is the
spline minimum. Consequently, the corrected time $t'_n$ for the $n$-th data
point can be  calculated as
$$t'_n =  t'_{n-1}+ \Delta t_n \left[{\int_{t_{n-1}}^{t_n} {s(t)\,{\rm
d}t}} \over {\Delta t_n s_{\rm min}} \right] ^{1/3}\,,$$  where $\Delta
t_n=t_n - t_{n-1}$ is the observed time interval between the $(n-1)$-th and
$n$-th data points.

By applying the above time correction to the $R$-band light curve, 
the total duration of the observing period
dilates from 241.71 to $285.57\rm\,d$.

As for the $R$-band flux autocorrelation, time correction does not
introduce any significant change in the results: the $\sim 7\rm\,h$
characteristic time scale of variability is confirmed, and no other signal
comes out.

Finally, DCF analysis has been applied to search for the possible existence of
time delays between the flux variations in the $B$ and $R$ bands: no
significant measurable (greater than a few minutes) delay has been found.

\section{Discussion and conclusions}

15625 ($=5^6$) observations were
performed in the period May 2000 -- January 2001 by 
24 telescopes of the WEBT collaboration to monitor the optical variability of BL
Lacertae. 

An exceptional sampling was reached, especially during the core WEBT
campaign (July 17 -- August 11, 2000), where time gaps were limited to a few
hours and they were essentially due to the lack of observers in the Pacific
area.

Well-defined intranight trends were detected, with variations up to $0.4\rm\,mag$ in
less than $2\rm\,h$.

Colour index analysis performed after subtracting the host galaxy
contribution from fluxes suggests the existence of at least two variability
mechanisms: the first one is essentially achromatic, and is
responsible for the long-term trend, while the second one, causing the
fast flares superposed on the long-term trend, implies spectral changes, the
spectrum becoming flatter when the source gets brighter. A similar behaviour was already 
found in the BL Lac object \object{S5 0716+71} by Ghisellini et al.\ (\cite{ghi97}).

As mentioned in the Introduction, Clements \& Carini (\cite{cle01}) argued
that the bluer-when-brighter behaviour might be due to a greater host galaxy
contribution when the AGN is fainter. Since we subtracted the host galaxy
contribution and still found the bluer-when-brighter trend (and only for
short-term variations), we conclude that spectral changes are not related to
the host galaxy contribution, but are an intrinsic property of fast flares.

The behaviour of the achromatic component determining the flux base-level 
modulations can be interpreted in terms of variation of the relativistic Doppler
factor $\delta$. By assuming that fluxes are enhanced proportionally to
$\delta^3$ by relativistic boosting, a maximum variation of $\delta$ of a
factor 1.36 is inferred. This change in $\delta$ can be due to
either a viewing angle variation of few degrees ($\Delta \theta \sim
1$--$4\degr$) or a noticeable change of the bulk Lorentz factor (more than
50\%). From  this point of view our interpretation would suggest that the flux
base-level modulations are more likely explained by a geometrical effect than
by an energetic one.

In a geometrical scenario such as that proposed by Villata \& Raiteri
(\cite{vil99}), where viewing angle variation is due to rotation of an
inhomogeneous helical jet, a rate of few
degrees (say,  1--$3\degr$) in a month (as can be inferred from the steepest
base-level  variations, see Fig.\ \ref{spline}) would imply an upper limit to
a possible periodicity of 10--$30\rm\,yr$. Indeed, in that model, a perfect
helix would give rise to a perfect periodicity of well-defined outbursts; if
the helical path presents distortions, the outburst phase (which can last for a
significant fraction of the period) is consequently disturbed by modulated
events whose steepest variations represent a lower limit to the rotation
rate.

We performed discrete autocorrelation analysis in
order to search for the existence of intermediate-short characteristic time
scales in our observing period. Variability on a typical time 
scale of $\sim 7\rm\,h$ was found
during the core WEBT campaign.

Cross-correlation analysis on the $B$ and $R$ fluxes does not reveal any 
significant time lag between variations in the two bands.

\begin{acknowledgements}
This research has made use of
the NASA/IPAC Extragalactic Database (NED), which is operated by the 
Jet Propulsion Laboratory, California Institute of Technology, under
contract with the National Aeronautics and Space Administration. 
This work was partly supported by the Italian Ministry for University and
Research (MURST) under grant Cofin 2001/028773 and by the Italian Space 
Agency (ASI) under contract CNR-ASI 1/R/27/00. It is
based partly on observations made with the Nordic Optical 
Telescope, operated on the island of La Palma jointly by Denmark, Finland, 
Iceland, Norway and Sweden in the Spanish Observatorio del Roque de 
Los Muchachos of the Instituto de Astrof\'{\i}sica de Canarias.  
St.-Petersburg group was supported by the Federal Program ``Integration''
under grants K0232 and A0007. J.\ Basler acknowledges support by the NASA
Missouri Space Grant Consortium.
J.\ H.\ Fan's work is partially supported by NSFC (19973001).

\end{acknowledgements}


\begin{thebibliography}{1999}
\bibitem[1969]{ber69} Bertaud Ch., Dumortier B., V\'eron P., et al., 1969,
A\&A 3, 436 
\bibitem[1979]{bes79} Bessell M.S., 1979, PASP 91, 589
\bibitem[1997]{blo97} Bloom S.D., Bertsch D.L., Hartman R.C., et
al., 1997, ApJ 490, L145 
\bibitem[2000]{boe00} B\"ottcher M., Bloom S.D., 2000, AJ 119, 469
\bibitem[2002]{boe02} B\"ottcher M., et al., 2002 (in preparation)
\bibitem[1989]{car89} Cardelli J.A., Clayton G.C., Mathis J.S., 1989, ApJ 345,
245
\bibitem[1992]{car92} Carini M.T., Miller H.R., Noble J.C.,
Goodrich B.D., 1992, AJ 104, 15
\bibitem[2001]{cle01} Clements S.D., Carini M.T., 2001, AJ 121, 90
\bibitem[2000]{cor00} Corbett E.A., Robinson A., Axon D.J., Hough J.H., 2000,
MNRAS 311, 485
\bibitem[2000]{den00} Denn G.R., Mutel R.L., Marscher A.P., 2000, ApJS 129, 61
\bibitem[1996]{dre96} Dreissigacker O., Camenzind M., 1996, PASPC 110, 377
\bibitem[1988]{ede88} Edelson R.A., Krolik J.H., 1988, ApJ 333, 646
\bibitem[1998]{fan98} Fan J.H., Xie G.Z., Pecontal E., Pecontal A., Copin Y.,
1998, ApJ 507, 173 
\bibitem[2001]{fan01} Fan J.H., Qian B.C., Tao J., 2001, A\&A 369, 758
\bibitem[1996]{fio96} Fiorucci M., Tosti G., 1996, A\&AS 116, 403
\bibitem[1997]{ghi97} Ghisellini G., Villata M., Raiteri C.M., et al., 1997, 
A\&A 327, 61
\bibitem[2000]{gho00} Ghosh K.K., Ramsey B.D., Sadun A.C., Soundararajaperumal
S., Wang J., 2000, ApJ 537, 638 
\bibitem[1997]{gro97} Grove J.E., Johnson W.N., 1997, IAU Circ.\ 6705
\bibitem[1997]{hag97} Hagen-Thorn V.A., Marchenko S.G., Mikolaichuk O.V.,
Yakovleva V.A., 1997, Astronomy Reports 41, 154 
\bibitem[2002]{hag02} Hagen-Thorn V.A., Larionov V.M., Hagen-Thorn A.V.,
Jorstad S.G., Temnov G.O., 2002, PASPC (in press)
\bibitem[1997]{har97} Hartman R.,
Bertsch D., Bloom S., Sreekumar P., Thompson D., 1997, IAU Circ.\ 6703 
\bibitem[1992]{huf92} Hufnagel B.R., Bregman J.N., 1992, ApJ 386, 473
\bibitem[1997]{mad97} Madejski G., Jaffe T., Sikora M., 1997, IAU Circ.\ 6705
\bibitem[1999]{mad99} Madejski G.M., Sikora M., Jaffe T., et al., 1999, ApJ
521, 145 
\bibitem[1997]{mae97} Maesano M., Montagni F.,
Massaro E., Nesci R., 1997, A\&AS 122, 267 
\bibitem[1997]{mak97} Makino F., Mattox J., Takahashi T., Kataoka J., 1997,
IAU Circ.\ 6708 
\bibitem[2001]{man01} Mannucci F., Basile F., Poggianti B.M., et al., 2001,
MNRAS 326, 745 
\bibitem[1996]{mar96} Marchenko S.G., Hagen-Thorn V.A.,
Yakovleva V.A., Mikolaichuk O.V., 1996, PASPC 110, 105
\bibitem[1996]{mars96} Marscher A.P., 1996, PASPC 110, 248
\bibitem[1998]{mas98} Massaro E., Nesci R., Maesano M., Montagni F., D'Alessio
F., 1998, MNRAS 299, 47 
\bibitem[1999]{mas99} Massaro E., Catalano S., Frasca A., et al., 1999,
in: Raiteri C.M., Villata M., Takalo L.O.\ (eds.) Proc.\ OJ-94 Annual
Meeting 1999, Blazar Monitoring towards the Third Millennium. Osservatorio
Astronomico di Torino, Pino Torinese, p.\ 10
\bibitem[1999]{mat99} Matsumoto K., Kato T., Nogami D.,
et al., 1999, PASJ 51, 253 
\bibitem[1999]{matt99} Mattox J.R., 1999, IAU Circ.\ 7189
\bibitem[1999]{mil99} Miller H.R., 1999, BAAS 31, 1508
\bibitem[1989]{mil89} Miller H.R., Carini M.T., Goodrich B.D., 1989, Nat 337,
627 
\bibitem[1977]{mil77} Miller J.S., Hawley S.A., 1977, ApJ 212, L47  
\bibitem[1978]{mil78} Miller J.S., French H.B., Hawley S.A., 1978, ApJ
219, L85 
\bibitem[1998]{nes98} Nesci R., Maesano M., Massaro E., et al.,
1998, A\&A 332, L1 
\bibitem[1999]{nik99} Nikolashvili M.G., Kurtanidze O.M., Richter G.M.,
1999, in: Raiteri C.M., Villata M., Takalo L.O.\ (eds.) Proc.\ OJ-94 Annual
Meeting 1999, Blazar Monitoring towards the Third Millennium. Osservatorio
Astronomico di Torino, Pino Torinese, p.\ 36
\bibitem[1997]{nob97} Noble J.C., Carini M.T., Miller H.R., et al., 1997, IAU
Circ.\ 6693 
\bibitem[2001]{pet01}Peterson B.M., 2001, in: The Starburst-AGN Connection 2001.
World Scientific, Singapore 
\bibitem[1992] {pre92} Press W.H., Teukolsky S.A., Vetterling W.T., Flannery
B.P., 1992, Numerical Recipes in Fortran - The Art of Scientific Computing.
Cambridge University Press, Cambridge
\bibitem[1970]{rac70} Racine R., 1970, ApJ 159, L99
\bibitem[2001]{rai01} Raiteri C.M., Villata M., Aller H.D., et al., 2001, 
A\&A 377, 396
\bibitem[2002]{rav02} Ravasio M., Tagliaferri G., Ghisellini G., et al., 2002, 
A\&A 383, 763
\bibitem[2000]{sca00} Scarpa R., Urry C.M., Falomo R., Pesce J.E., Treves A.,
2000, ApJ 532, 740
\bibitem[1987]{sch87} Schneider P., Weiss A., 1987, A\&A
171, 49 
\bibitem[1995]{smi95} Smith A.G., Nair A.D., 1995, PASP 107, 863
\bibitem[1999]{sob99} Sobrito G., Villata M., Raiteri C.M., et al., 1999,
Blazar Data 1, 5 
\bibitem[1998]{spe98} Speziali R., Natali G., 1998, A\&A 339,
382 
\bibitem[2000]{tan00} Tanihata C., Takahashi T., Kataoka J., et al., 2000,
ApJ 543, 124 
\bibitem[1999]{tos99} Tosti G., Luciani M., Fiorucci M., et al., 1999, Blazar
Data 2, 1 
\bibitem[1995]{ver95} Vermeulen R.C., Ogle P.M., Tran H.D., et
al., 1995, ApJ 452, L5
\bibitem[1999]{vil99} Villata M., Raiteri C.M., 1999, A\&A 347, 30
\bibitem[2000]{vil00} Villata M., Mattox J.R., Massaro E., et al., 2000, A\&A
363, 108 
\bibitem[2002]{vil02} Villata M., et al., 2002, A\&A (in preparation) 
\bibitem[1988]{web88} Webb J.R., Smith A.G.,
Leacock R.J., et al., 1988, AJ 95, 374 
\bibitem[1998]{web98} Webb J.R., Freedman I., Howard E., et al., 1998, AJ 115,
2244
\bibitem[1996]{wii96} Wiita P.J., 1996, PASPC 110, 42

\end{thebibliography}
\end{document}